\documentclass[doublecol,amsmath,amssymb]{epl2}
\usepackage{graphicx}
\usepackage{dcolumn}
\usepackage[dvipsnames]{xcolor}

\begin{document}

\title{Stability in Chaos}

\author{Greg Huber$^{1,2}$, Marc Pradas$^{3,1}$, 
Alain Pumir$^{4,1}$ and Michael Wilkinson$^{3,1}$}

\institute{
$^1$ Kavli Institute for Theoretical Physics, Kohn Hall, 
University of California, Santa Barbara, CA93106, USA\\
$^2$ Department of Physics, University of California, Santa Barbara, CA93106, USA\\
$^3$ School of Mathematics and Statistics,
The Open University, Walton Hall, Milton Keynes, MK7 6AA, UK\\
$^4$ Univ Lyon, ENS de Lyon, Univ Claude Bernard, CNRS, 
Laboratoire de Physique, F-69342, Lyon, France\\
}

\abstract{
Intrinsic instability of trajectories characterizes chaotic dynamical systems.
We report here that trajectories can exhibit
a surprisingly high degree of stability, over a very long time, 
in a chaotic dynamical system. We 
provide a detailed quantitative
description of  this effect for a one-dimensional model of inertial
particles in a turbulent flow using large-deviation theory. 
Specifically, the determination of the entropy function for the 
distribution of finite-time Lyapunov exponents reduces to the analysis of a
Schr\"odinger equation, which is tackled by semi-classical methods.
}


\maketitle

\section{Introduction}
\begin{figure}[h!]
\includegraphics[width=0.48\textwidth]{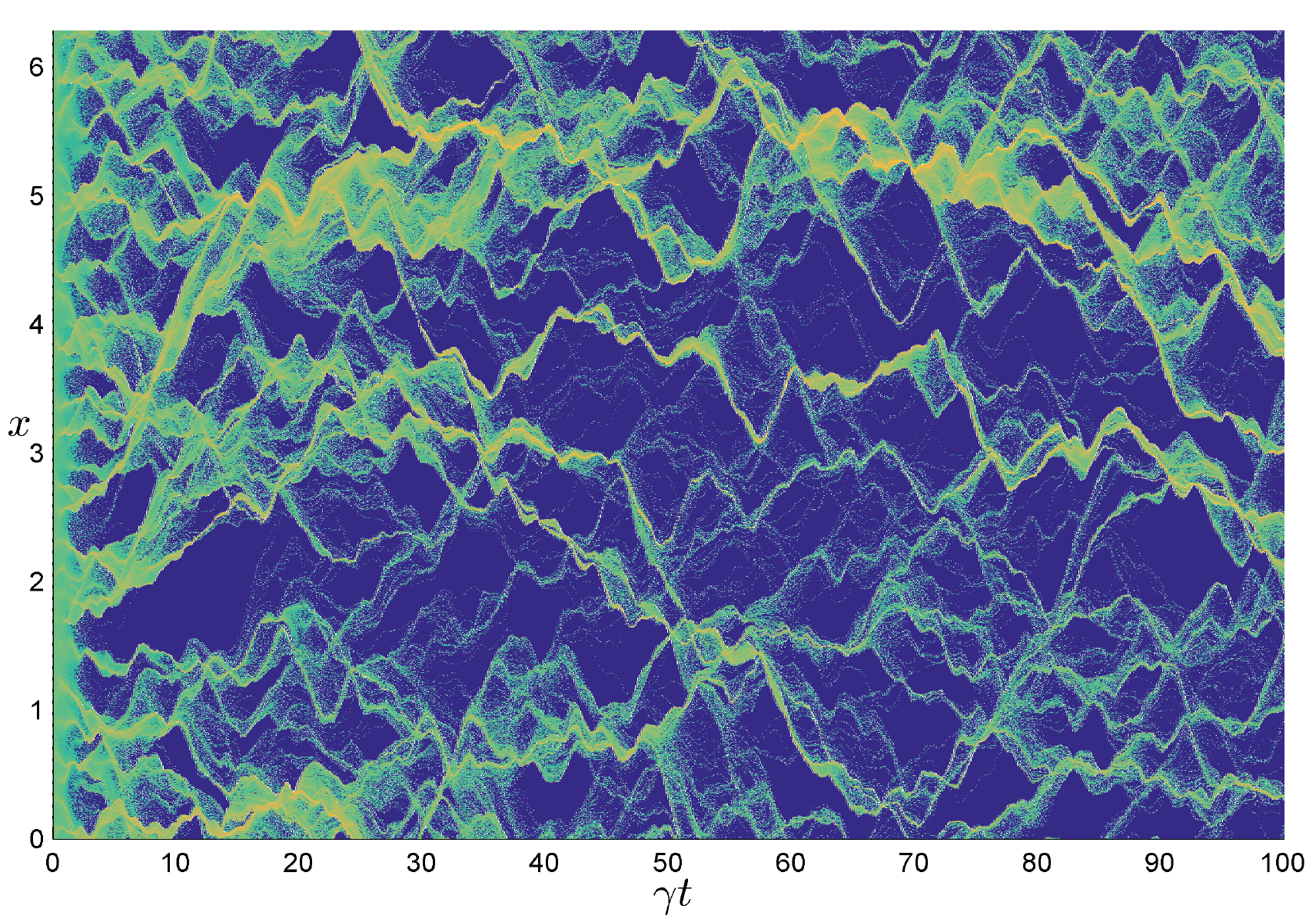}
\caption{ %(Color online.) 
Trajectories, $x(t)$, for the dynamical system described by 
Eq.~(\ref{eq: 3}) with $\xi=0.08$, $L=2\pi$, and the 
dimensionless parameter [cf.~Eq.~(\ref{eq: 11})] is $\epsilon=1.7678$.
}
\label{fig: 1}
\end{figure}

This Letter concerns a phenomenon illustrated by the peculiar nature
of the trajectories $x(t)$ of inertial particles (Fig.~\ref{fig: 1}) in a 
one-dimensional model, which is 
described in detail later (Eq.~(\ref{eq: 3})). The plot 
shows a very large number of trajectories, which start 
with a uniform initial density. The trajectories clearly show a strong
tendency to cluster, and the plot (online version) is color-coded using a logarithmic 
density scale to illustrate the very intense accumulation of probability density 
in some regions. Clustering of trajectories of a dynamical system is usually 
characterised by showing that the highest Lyapunov exponent 
of the dynamics is negative \cite{Ott02}, and conversely a positive Lyapunov exponent 
is the essential characteristic of chaotic dynamics. The flow illustrated in 
Fig.~\ref{fig: 1}, however, is known to have a positive Lyapunov exponent, so 
the very marked clustering is only transient, as trajectories must
eventually separate exponentially. 

Earlier work has shown that one-dimensional chaotic systems may 
exhibit 
a temporary convergence preceding their eventual separation  
(see, e.g.~\cite{Fuj83, Aur+96}), and it has been argued that 
the predictability of dynamical systems can be very strongly dependent 
on initial conditions \cite{Smi+99,Smi94}. Figure \ref{fig: 1}, however, 
reveals that: (a) the convergence can lead to clusters of trajectories over 
times 
which are much longer than the expected divergence time, and (b)  the simulated 
trajectories tend to form surprisingly dense clusters. 
It is the principal objective of this Letter to 
describe and quantify the extent to which the phase space of this chaotic system 
is permeated by islands of transient stability, and to argue that the 
reasoning extends to typical chaotic systems.
It complements another work \cite{Pra+17} 
which quantifies the intensity of the clustering effect, and which also
shows examples of similar clustering effects in other dynamical systems. 
In the concluding remarks, we argue that this 
phenomenon may be applicable to pricing futures and insurance contracts.

\section{Distribution of sensitivity to initial conditions}
\begin{figure}[t!]
\includegraphics[width=0.48\textwidth]{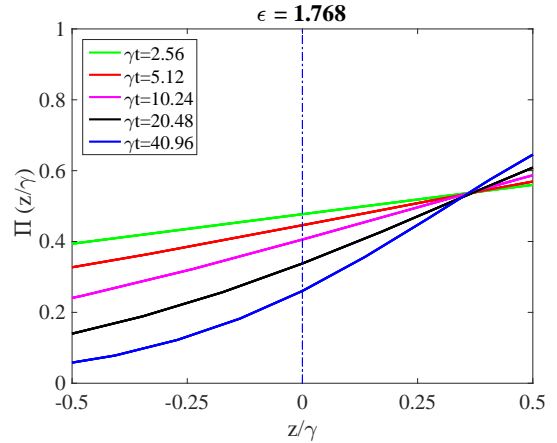}
\caption{%(Color online.) 
Cumulative probability, $\Pi$, for the value of the FTLE, $z(t)$,  at different  
times (in dimensionless units). The distribution of $z(t)$ is very broad, even for large values 
of $t$. The parameters are the same as for Fig.~\ref{fig: 1}.
}
\label{fig: 2}
\end{figure}

The tendency of the trajectories to exhibit converging behavior is illustrated in 
Fig.~\ref{fig: 2}, which shows the cumulative probability, $\Pi$,
for the finite-time Lyapunov exponent (FTLE) at long times. The FTLE at time $t$ for a 
trajectory starting at $x_0$ is defined by 
\begin{equation}
\label{eq: 1}
z(t)=\frac{1}{t}\ln\,\left\vert\frac{\partial x_t}{\partial x_0}\right\vert_{x(0)=x_0},
\end{equation}
where $x_t$ denotes position at time $t$. The expectation value of $z(t)$ in the limit 
as $t\to \infty$ is termed the Lyapunov 
exponent: $\Lambda= \lim_{t \rightarrow \infty} \langle z(t)\rangle$ (angular brackets denote 
ensemble  averages 
throughout). When $\Lambda>0$, there is an almost 
certain exponential growth of infinitesimal separations of trajectories. For the 
example in Fig.~\ref{fig: 1}, we have $\Lambda=0.075\,\gamma$, where 
$\gamma$ is a positive parameter of the model [cf.~Eq.~(\ref{eq: 3})]. 
Figure \ref{fig: 2} shows that 
the cumulative probability distribution for $z$ is very broad: 
even at time $t=41/\gamma$, 
which is comparable to the duration of the trajectories shown in Fig.~\ref{fig: 1}, 
the probability of $z$ being negative is as high as $0.25$. We shall see how this very 
broad distribution can be quantified.

It is usually assumed that when the highest Lyapunov 
exponent is positive, the long-term behavior of a system is inherently 
unpredictable because of exponential sensitivity to the initial conditions. 
However, the phenomenon illustrated in Figs.~\ref{fig: 1} and \ref{fig: 2}
indicates that there may be basins in the space of initial conditions which attract a significant 
fraction of the phase space, giving a final position which is highly insensitive 
to the initial conditions. If the initial conditions which are of physical interest lie 
within one of these basins, the behavior of the system can be computed 
accurately for a time which is many multiples of the inverse of the Lyapunov 
coefficient. 

Next we describe the equations of motion which were used to generate Fig.~\ref{fig: 1}. 
They correspond to
\begin{eqnarray}
\label{eq: 3}
\dot x&=&v,
\nonumber \\
\dot v&=&\gamma[u(x,t)-v],
\end{eqnarray}
where  $x$ and  $v$ are the  position and velocity, respectively, of a small particle in a viscous fluid
\cite{Gat83,Max+83}; $\gamma$ is a constant describing the rate of damping 
of the motion of a small particle relative to the fluid and $u(x,t)$ 
is a randomly fluctuating velocity field of the fluid in which the 
particles are suspended. In Fig.~\ref{fig: 1} we simulated a velocity 
field where the correlation function is white noise in time, satisfying
$\langle u(x,t)\rangle=0$ and 
$\langle u(x,t)u(x',t')\rangle =\delta(t-t')C(x-x')$.  
The correlation function is 
$C(\Delta x)={\cal D}\xi^2\,\exp\left(-\Delta x^2/2\xi^2\right)$, 
where ${\cal D}$ and $\xi$ are constants. 
Trajectories which leave the interval $[0,L]$ are returned there
by adding a multiple of $L$ to $x$. 
Equation (\ref{eq: 3}) and related models have been studied 
intensively as descriptions of particles suspended in turbulent flows: see  
\cite{Fal+01} and \cite{Gus+16} for reviews.

\section{Large-deviation analysis}
\begin{figure}[t!]
\includegraphics[width=0.48\textwidth]{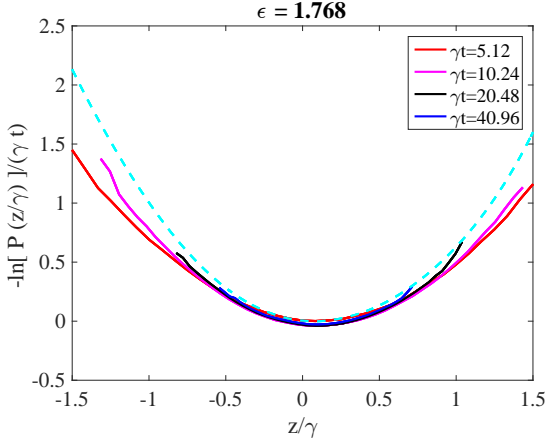}
\caption{%(Color online.) 
The transformed probability density function $-\ln\,P(z)/t$ approaches 
a limit, termed the large deviation entropy function $J(z)$. 
When $t \to \infty$, we find excellent agreement with a 
theoretical prediction for $J(z)$ (dashed line).
}
\label{fig: 3}
\end{figure}

In the large-time limit the probability 
density of $z$ is expected to be described by a large deviation 
approximation \cite{Fre+84,Tou09}:
\begin{equation}
\label{eq: 2}
P(z)\sim \exp[-tJ(z)],
\end{equation}
where $J(z)$ is termed the \emph{entropy function} or the \emph{rate function}. 
Large deviation methods have previously been applied to analyse the 
distribution of finite-time Lyapunov exponents in a variety of contexts: \cite{Tan+03}
and \cite{Tai+07} are representative examples. In this Letter we
are able to explain the broad distribution illustrated in Fig.~\ref{fig: 2} by determining 
the entropy function $J(z)$: if the second derivative, $J''(\Lambda)$, is small, the FTLE 
has a very broad distribution, giving a quantitative explanation for Fig.~\ref{fig: 2}.  
In Fig. \ref{fig: 3} we transform the empirical distributions of $z$ for different
values of the time $t$ to determine the entropy function $J(z)$: the fact that the curves 
for different values of $t$ are quite accurately superimposed implies that the values 
of $t$ displayed in Fig. \ref{fig: 2} are already sufficiently large for large deviation 
theory to be applicable. In Fig.~\ref{fig: 3} we also compare the entropy function 
obtained from our empirical distributions of $z$ with a theoretical curve (described below). 
There is very satisfactory agreement as $t\to \infty$, indicating that the effect
illustrated in Figs.~\ref{fig: 1} and \ref{fig: 2} has been understood quantitatively.

Our theoretical approach involves the analysis of a cumulant, $\lambda(k)$, which is 
defined by
\begin{equation}
\label{eq: 6}
\langle \exp(kzt)\rangle =\exp\left[t\lambda(k)\right].
\end{equation}
The large deviation principle, as represented 
by equation (\ref{eq: 2}), implies that 
\begin{equation}
\label{eq: 7}
\langle \exp(kzt)\rangle=\int_{-\infty}^\infty{\rm d}z\ \exp[t(kz-J(z))].
\end{equation}
A Laplace estimate shows that $\lambda$ and $J$ are a Legendre transform 
pair:
\begin{equation}
\label{eq: 8}
\lambda(k)=kz-J(z), \ \ \  J'(z)=k. 
\end{equation}
For the model described by Eq.~(\ref{eq: 3}), the cumulant can be 
determined as an eigenvalue of a differential equation. 
Following the approach discussed in \cite{Wil+15}, 
we can obtain a Fokker-Planck equation for the variables $Y$ and $Z$ defined by 
$Z=\frac{\delta \dot x}{\delta x}$ and $Z=\dot Y$:
\begin{equation}
\label{eq: 9}
\frac{\partial \rho}{\partial t}=
-\partial_ Y(Z\rho)+\hat {\cal F}\rho, 
\end{equation}
where we have defined $\hat {\cal F}\rho\equiv \partial_Z(v(Z)\rho)+{\cal D}\gamma^2 \partial_Z^2\rho$
with $v(Z)=-\gamma Z-Z^2$. Note that $Y=zt$, and we introduce the Lyapunov
exponent, $\Lambda=\langle Z\rangle$. The cumulant $\lambda(k)$ is the largest 
eigenvalue of the operator $\hat {\cal F} + k Z$ \cite{Don+76}:
\begin{equation}
\label{eq: 10}
\hat {\cal F}\rho(Z)+kZ\rho(Z)=\lambda(k)\rho(Z)
\ .
\end{equation}
It is convenient to make a transformation of coordinates:
\begin{equation}
\label{eq: 11}
x=\left(\gamma {\cal D}\right)^{-1/2}Z
\ ,\ \ \ 
\epsilon=\sqrt{\frac{{\cal D}}{\gamma}}
\ ,\ \ \ 
E=-\frac{\lambda}{\gamma}
\ .
\end{equation}
The parameter $\epsilon$ is a dimensionless measure of the strength 
of inertial effects in the model (\ref{eq: 3}). It is known 
that the Lyapunov exponent $\Lambda $ is negative, indicating almost 
certain coalescence of paths, when $\epsilon<\epsilon_{\rm c}=1.3309\ldots$ 
\cite{Wil+03}. For $\epsilon>\epsilon_{\rm c}$, the Lyapunov exponent is positive 
so that the motion is chaotic. All of the illustrations in this paper are at 
$\epsilon=1.7678 \approx 1.33\ldots\times\ \epsilon_{\rm c}$, where $\Lambda\approx 0.075 \ \gamma$.  
In the coordinates defined by (\ref{eq: 11}), the cumulant obeys an equation of the form
\begin{equation}
\label{eq: 12}
\partial_x(\partial_x+x+\epsilon x^2)\rho(x)+k\epsilon x\rho(x)+E \rho(x)=0
\ .
\end{equation}

\section{WKB method for cumulants} 
We now transform (\ref{eq: 12}) so that it takes the form of a 
Schr\"odinger equation. Write $\hat {\cal F}=\partial_x[\partial_x+x+\epsilon x^2]$ and consider 
a transformation $\hat {\cal H}=\exp[-\Phi(x)]\hat {\cal F}\exp[\Phi(x)]$ with
$ \Phi(x)=-x^2/4-\epsilon x^3/6$. The cumulant $\lambda(k)$ is then obtained from the 
ground-state eigenvalue $E_0$ of a Hermitean operator
\begin{equation}
\label{eq: 14}
\psi''-V(x)\psi =E \psi
\end{equation}
where $\lambda=-E_0/\gamma$ and the potential is 
\begin{equation}
\label{eq: 15}
V(x)=\frac{1}{4}(x+\epsilon x^2)^2-\frac{1}{2}-\epsilon(k+1)x
\ .
\end{equation}
Note that Eq.~(\ref{eq: 14}) corresponds to a Schr\"odinger equation with $m=\frac{1}{2}$ and $\hbar=1$. 
We remark that, when $\epsilon$ is small, the potential $V(x)$ has two minima, 
close to $x=0$ and to $x=-1/\epsilon$.

The WKB method \cite{Hea62,Lan+58} provides a powerful tool for understanding the structure 
of solutions of the Schr\"odinger equation. It works best when the potential energy is slowly 
varying. In the case of equation (\ref{eq: 14}), $\epsilon$ is the small 
parameter of WKB theory, because the minima of the potential move apart as 
$\epsilon\to 0$. In fact, a change of variable $x = \epsilon X$ formally
reduces Eq.~(\ref{eq: 14}) to an expression where the $\psi''$ term has a small 
coefficient.
We will find, however, that WKB methods yield surprisingly accurate results 
even when $\epsilon$ is not small. Define the momentum
\begin{equation}
\label{eq: 16}
p(x)=+\sqrt{V(x)-E},
\end{equation}
with $p(x)=0$ where $V(x)<E$. The action integral is
\begin{equation}
\label{eq: 17}
S(x)=\int_0^x{\rm d}x'\ p(x'),
\end{equation}
and define a pair of WKB functions
\begin{equation}
\label{eq: 18}
\phi_\pm(x)=\frac{1}{\sqrt{p(x)}}\exp[\pm S(x)]
\ .
\end{equation}
Then, as we get further away from turning points where $p(x)=0$, the solutions 
of (\ref{eq: 14}) are asymptotic to a linear combination of WKB solutions 
$f(x)=a_+ \phi_+(x)+a_- \phi_-(x)$, 
where $a_\pm$ are approximately constant, except in the vicinity of turning points
where $E=V(x)$. 

The Schr\"odinger equation (\ref{eq: 14}) has unusual boundary conditions.
The gauge transformation implies that the solutions of (\ref{eq: 12}) 
and (\ref{eq: 14}) are related by
\begin{equation}
\label{eq: 20}
\psi(x)=\exp[-\Phi(x)]\rho(x)=\exp\left[\frac{x^2}{4}+\epsilon\frac{x^3}{6}\right]\rho(x)
\ .
\end{equation}
Integration of equation (\ref{eq: 12}) gives 
\begin{equation}
\label{eq: 21}
\int_{-\infty}^\infty {\rm d}x\ (k\epsilon x+E)\rho(x)=0,
\end{equation}
so that $\langle x\rangle$ must be finite. Then equation (\ref{eq: 20}) implies that 
the coefficient of $a_-$ must be zero as $x\to -\infty$ (so that $\rho(x)$ does not 
diverge). Furthermore, $\rho(x)$ has an algebraic decay as $x\to \pm \infty$ and 
the coefficients of these algebraic tails must be equal in order for $\langle x\rangle$ 
to be finite. In terms of the coefficients $a_\pm$, the appropriate boundary 
conditions are therefore $\lim_{x\to -\infty}a_+=1$ and $\lim_{x\to -\infty}a_-=0$. 
At large positive values of $x$, we have
\begin{equation}
\label{eq: 22}
\lim_{x\to +\infty}\left\{
\begin{array}{ll}
a_+  &=\exp(\Sigma)  \\ 
a_-  &=c
\end{array}\right.
\end{equation}
where $c$ is indeterminate, and where $\Sigma$ is defined by the 
finite limit of the following expression:
\begin{equation}
\label{eq: 23}
\Sigma=\lim_{x\to \infty}\left[S(x)-S(-x)-\Phi(x)+\Phi(-x)\right]
\ .
\end{equation}
We can determine the smallest eigenvalue $E(k)$ by using a shooting method
to find a solution which satisfies (\ref{eq: 22}).  Solving numerically (\ref{eq: 14}) amounts 
to propagating a two-dimensional vector 
$\mbox{\boldmath$a$}(x)=(\psi(x),\psi'(x))$. We can take an 
initial condition for $x_{\rm i}$ large and negative in the form
$\mbox{\boldmath$a$}_{\rm i}=\left(1,p(x_{\rm i})\right)\exp(\Phi(x_{\rm i}))$, 
corresponding to the asymptotic form of the solution which decays as 
$x\to -\infty$.  We numerically propagate this solution for increasing $x$, and find
that the solution increases exponentially. If the first element of the solution vector at
$x_{\rm f}\gg 1$ is $a_1(x_{\rm f})=\psi(x_{\rm f})$, we can express 
the eigenvalue condition in the following form:
\begin{equation}
\label{eq: 24}
f(k,\epsilon,E)\equiv \frac{\psi(x_{\rm f})\exp[\Phi(x_{\rm f})]}
{\psi(x_{\rm i})\exp[\Phi(x_{\rm i})]}=1
\ .
\end{equation}
This shooting method does  provide very accurate values for the 
cumulant $\lambda=-\gamma E_0$. We used this method to 
obtain the cumulant. Performing a Legendre transform gives the 
theoretical curve in Figure \ref{fig: 3}. We remark that while the 
entropy function is well approximated by a quadratic, corresponding 
to the FTLE having an approximately Gaussian distribution for the
parameter values reported here, our calculation can be used to accurately 
determine the non-Gaussian tails of the distribution of the FTLE.

\section{Bohr-Sommerfeld quantisation for cumulant}
It is also desirable to be able to make analytical estimates of the eigenvalues.  
The coefficients $a_\pm$ can be approximated as changing discontinuously 
when $x$ passes a turning point, where $E-V(x)$ is zero (or close to zero). 
Depending on the value of $E$ there may be one or two double turning 
points. We must take account of the fact that the amplitudes $a_\pm$ can 
change \lq discontinuously' in the vicinity of turning points. 
Close to a double turning point, the equation is approximated by a parabolic 
cylinder equation
\begin{equation}
\label{eq: 25}
\frac{{\rm d}^2\psi}{{\rm d}x^2}-\frac{1}{4}x^2\psi+E \psi=0.
\end{equation}
We are interested in constructing a solution which is exponentially increasing 
as $x$ increases, both when $x\to -\infty$ and for $x\to +\infty$. We can use this solution 
in the form $\phi(x)=A(x)\,\exp[S(x)]/\sqrt{p(x)}$ 
where $A(x)$ is asymptotically constant as $x\to\pm \infty$, and we take
$A(-\infty)=1$. By adapting a calculation due to Miller and Good \cite{Mil+53}, 
we find that as $x\to +\infty$, the solution is approximated by $A(x)=F(E)$, where  
he function $F(E)$ is 
\begin{equation}
\label{eq: 26}
F(E)=\frac{\sqrt{2\pi}}{\Gamma(\frac{1}{2}-E)}\exp[E(1-\ln\,|E|)],
\end{equation}
and has zeros at $E=n+\frac{1}{2}$, $n=0,1,2,\ldots$. 
It approaches unity as $E\to -\infty$ and 
it oscillates approximately sinusoidally with amplitude equal to $2$ as $E\to +\infty$. 
Equation (\ref{eq: 26}) can be used to determine the amplitude of the 
exponentially increasing solution after passing through a double turning point. 

\begin{figure}[t!]
\includegraphics[width=0.48\textwidth]{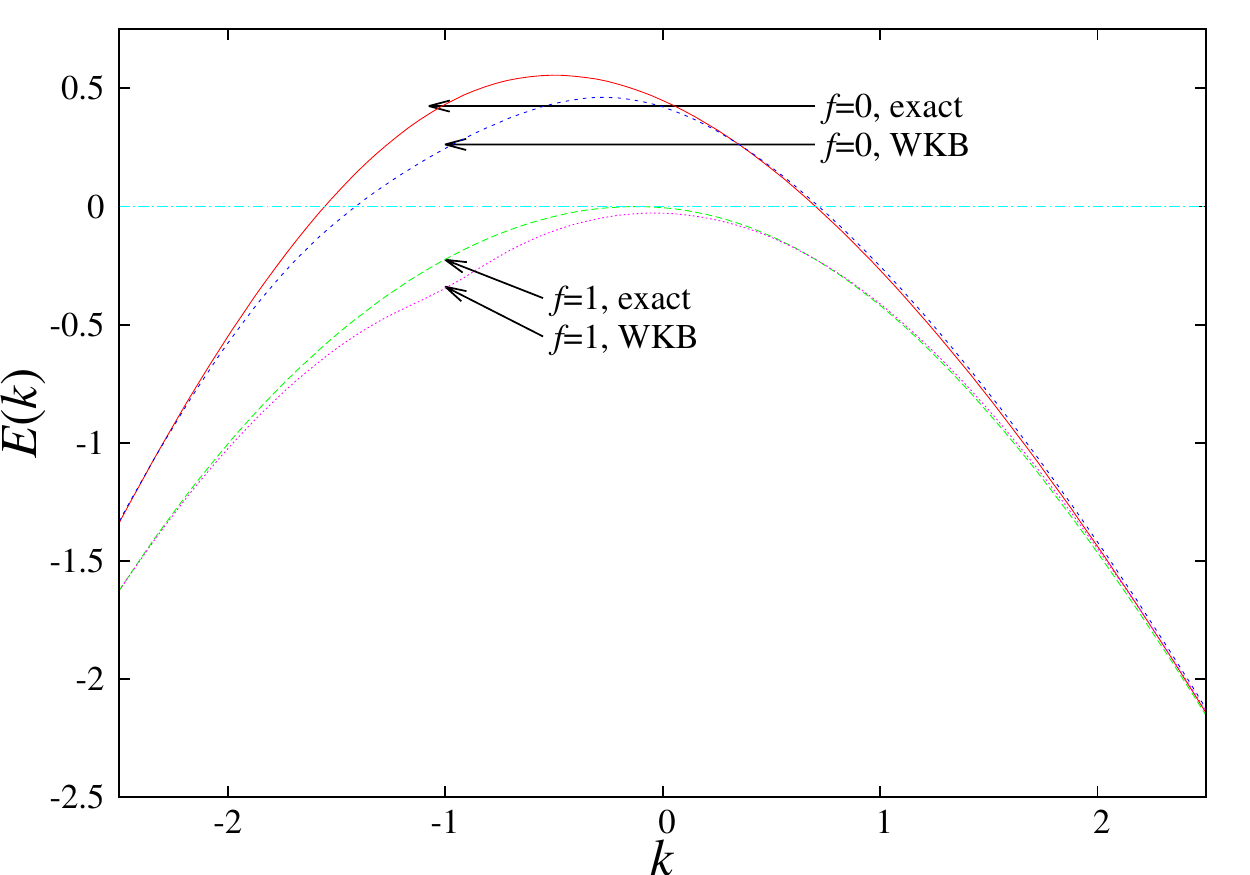}
\caption{%(Color online.) 
The generalised Bohr-Sommerfeld quantisation condition, Eq.~(\ref{eq: 28}), 
produces remarkably accurate eigenvalues. The upper curves are eigenvalues with 
the usual boundary condition (square-integrable wavefunction), with $f=0$, comparing the
numerically exact eigenvalue with that obtained by the Bohr-Sommerfeld condition. 
The lower curves are for the criterion $f=1$ which applies to our cumulant eigenvalues. 
These data are for the case $\epsilon =1.7678$, and the dashed line in Fig.~\ref{fig: 3}
is the Legendre transform of the numerically exact eigenvalue for $f=1$.
}
\label{fig: 4}
\end{figure}

The eigenvalue condition (\ref{eq: 24}) can also be expressed using the WKB 
approximation, leading to a generalisation of the Bohr-Sommerfeld quantisation 
condition. We consider cases where the potential has a closely spaced pair 
of real turning points, which will be treated as a double turning point, close to $x=0$. 
The effect of the double turning point is to cause the WKB amplitude 
of the exponentially increasing solution $\phi_+(x)$ to change 
by a factor $F(E)$, which we assume to be given correctly by the expression for a 
parabolic potential, Eq.~(\ref{eq: 26}). Because the potential is not 
precisely parabolic at the double turning point, the energy argument of $F(E)$ 
should be replaced by $F(\sigma/\pi)$, where $\sigma$ is a phase integral:
\begin{equation}
\label{eq: 27}
\sigma=\int_{x_1}^{x_2}{\rm d}x\ \sqrt{E-V(x)},
\end{equation}
with $x_1$ and $x_2$ being the turning points where $E=V(x)$. This 
is the most natural choice of replacement variable, because it reproduces 
the standard Bohr-Sommerfeld condition in the case where the solution of the 
Schr\"odinger equation is square-integrable. In the more general case that we
consider, the WKB eigenvalues are the solutions of 
\begin{equation}
\label{eq: 28}
F(\sigma/\pi)\exp(\Sigma)=f
\end{equation}
where $f=1$ corresponds to the correct boundary condition for our eigenvalue 
equation. Equation (\ref{eq: 28}) is a generalisation of the usual Bohr-Sommerfeld 
condition, and it corresponds with the standard form of the Bohr-Sommerfeld criterion, 
which applies to bound-state problems, when $f=0$. We find that 
it does produce remarkably accurate eigenvalues, as illustrated in Fig.~\ref{fig: 4}, despite 
that fact that $\epsilon$ is not small. In order to emphasise the 
fact that the modified Bohr-Sommerfeld condition does give very different 
eigenvalues, in Fig. \ref{fig: 4} we display results for the 
conventional Bohr-Sommerfeld condition, $f=0$, as well as for 
$f=1$, which approximates the cumulant. We see that the modified Bohr-Sommerfeld condition 
provides accurate information about the cumulant $\lambda (k)$ in terms of two 
integrals of the momentum $\sqrt{V(x)-E}$, namely $\Sigma$ and $\sigma$.

We remark that Fyodorov {\sl et al} have studied very closely related equations
which occur in modelling pinning of polymers, including a related WKB analysis \cite{Fyo+17}.

\section{Conclusions}
We have demonstrated, for the system described by Equations (\ref{eq: 3}),  
that the usual definition of chaos, 
based on the instability of trajectories in the long time limit, 
does not preclude the existence of large islands of long term stability 
illustrated by the clustering of trajectories in Fig.~\ref{fig: 1}.  
We argued that this clustering is related to the broad
distribution of finite-time Lyapunov exponents, with a large probability of
having negative values. In our analysis of Equations (\ref{eq: 3}) we 
determined the cumulant, and performed a Legendre transform to obtain the large-deviation 
entropy function of the FTLE. We further showed how Bohr-Sommerfeld quantisation 
gives an accurate approximation to the cumulant. This analytical approach allows considerable 
scope for generalisation, for example to determine analytical approximations to the correlation 
dimension which describes the clustering of trajectories \cite{Wil+12,Wil+15}. We expect to 
explore this in a subsequent publication. We also anticipate that the methods will find quite direct 
application to clustering of particles advected on fluid surfaces, such as is seen in experiments 
reported in \cite{Som+93} and \cite{Lar+09}. 

We should consider the extent to which the behaviour illustrated in Figure \ref{fig: 1}
is expected to be a general feature of chaotic dynamical systems. The differential 
structure of the equations 
of motion (\ref{eq: 3}) has no properties which distinguish it from a generic dynamical system, 
and our argument was based upon considering the distribution of the FTLE, which has 
generic properties. In fact we can propose a simple criterion for observing the effect illustrated
in Figures \ref{fig: 1} and \ref{fig: 2}. We showed that the clustering is a consequence of there
being a substantial probability to observe a negative FTLE at time $t$. 
Using Equation (\ref{eq: 2}), and making a quadratic approximation for 
$J$, we have $P(z)\sim \exp[-tJ''(z-\Lambda)^2/2]$, which 
indicates that $P(0)$ is of order unity up to a dimensionless timescale given by:
\begin{equation}
\label{eq: 29}
t^\ast\Lambda=\frac{2}{J''(\Lambda)\Lambda}
\ .
\end{equation}
The natural expectation is that transient clustering may occur on a 
timescale such that $\Lambda t$ is of order unity. However, equation (\ref{eq: 29}) 
indicates that the timescale over which transient clustering is observed may be much 
larger, and that $J''(\Lambda)/\Lambda$ is the relevant dimensionless measure of the clustering 
effect illustrated in Figure \ref{fig: 1}. This quantity diverges at the transition to chaos, and 
it may remain large even when the system is not close to a transition. For example, in Equations 
(\ref{eq: 3}) we have $2/J''\Lambda\approx 13$ when $\epsilon/\epsilon_{\rm c}=1.33$. 
We remark that the dimensionless parameter in Eq. (\ref{eq: 29}) can be expressed in terms 
of an integral of a correlation function:
\begin{equation}
\label{eq: 30}
t^\ast \Lambda=\frac{2}{\Lambda}\int_{-\infty}^\infty {\rm d}t\ \left[\langle Z(t)Z(0)\rangle-\Lambda^2\right]
\end{equation}
where $Z(t)=\frac{{\rm d}}{{\rm d}t}\left[tz(t)\right]$. This expression can be useful in cases, such 
as Eq. (\ref{eq: 3}), where it is practicable to write an equation of motion for $z(t)$ \cite{Wil+12}. 
It is readily derived by estimating the variance of $tz(t)$.

Smith and co-workers (see \cite{Smi94,Smi+99}, and references cited therein) 
have emphasised the wide variability of the local instability of chaotic dynamical systems, 
indicating that the Lyapunov exponent is not sufficient to characterise chaos.
Our work indicates that the transient stability can be very long-lived, and we 
propose that $1/J''\Lambda$ should be adopted as a parameter characterising the 
transient stability lifetime of chaotic systems. 
Our observation that trajectories of a generic chaotic 
system may be stable for surprisingly long times over a 
substantial domain of phase space implies in 
practice that small perturbations may not be amplified, making the system
``predictable" longer than naturally expected.
One potential application of this observation is to 
insurance or futures transactions, where someone
takes a fee in exchange for writing a contract which requires a payment 
to be made if there is a loss or an unfavourable change in the price. 
The predictability of the behavior of the system over very long times
for certain initial conditions, 
implied by our work, may be used to gain advantage. 
In some areas, such as weather-dependent risks, it may be 
possible to understand the conditions leading to a much smaller
uncertainty than expected, so that the risk in a contract would be 
reduced. Finally, we remark that there are relations between our results 
and studies of the possibility of \textit{negative} entropy production in systems out of
equilibrium~\cite{GC95, Sei12}. The two processes are different because entropy 
is a property of phase-space volume, whereas the Lyapunov exponent describes 
distances between phase points. Whether the theoretical results developed in
this context can lead to a better understanding of our
system remains to be explored.

The authors are grateful to the Kavli Institute for Theoretical Physics for 
support, where 
this research was  supported in part by the National Science Foundation 
under Grant No. PHY11-25915. We appreciate stimulating discussions with 
Arkady Vainshtein (on asymptotics and supersymmetry) and Robin Guichardaz 
(on large-deviation theory).


\begin{thebibliography}{}

\bibitem{Ott02}
\Name{Ott, E.}
\Book{Chaos in Dynamical Systems, 2nd edition}
\Publ{University Press, Cambridge}
\Year{2002}

\bibitem{Fuj83}
\Name{Fujisaka, H.}
%{\sl  Statistical dynamics generated by fluctuations of local Lyapunov exponents},
\REVIEW{Prog. Theor. Phys.}{70}{1983}{1264}

\bibitem{Aur+96}
\Name{E. Aurell, E., Boffetta, G., Crisanti, A., Paladin, G. and Vulpiani, A.}
%{\sl Growth of Non-infinitesimal Perturbations in Turbulence,}
\REVIEW{Phys. Rev. Lett.}{77}{1996}{1262}

\bibitem{Smi+99}
\Name{Smith, L. A., Ziehmann, C. and Fraedrich, K.}
%{\sl Uncertainty dynamics and predictability in chaotic dynamical systems}
\REVIEW{Q. R. J. Meteor. Soc.}{125}{1999}{2855-86}

\bibitem{Smi94}
\Name{Smith, L. A.}
%{\sl Local optimal prediction: exploiting strangeness and the variation of strangeness to initial condition}
\REVIEW{Phil. Trans. Roy. Soc.}{348}{1994}{371-81}

\bibitem{Pra+17}
\Name{Pradas, M., Pumir, A, Huber, G. and Wilkinson, M.} 
%{\sl Convergent Chaos},
\REVIEW{J Phys A}{50}{2017}{275101}

\bibitem{Gat83}
\Name{Gatignol, R.}
%{\sl Faxen formulae for a rigid particle in an unsteady non-uniform Stokes flow},
\REVIEW{J. M\'ec. Th\'eor. Appl.}{1}{1983}{143-60}

\bibitem{Max+83}
\Name{Maxey, M. R. and Riley, J. J.} 
%{\sl Equation of motion for a small rigid sphere in a nonuniform flow},
\REVIEW{Phys. Fluids}{26}{1983}{883-9}

\bibitem{Fal+01}
\Name{Falcovich, G., Gawedzki, K. and Vergassola, M.}
%{\sl Particles and fields in fluid turbulence}
\REVIEW{Rev. Mod. Phys.}{73}{2001}{913-75}

\bibitem{Gus+16}
\Name{Gustavsson, K and Mehlig, B.}
%{\sl Statistical models for spatial patterns of inertial particles in turbulence}
\REVIEW{Adv. Phys.}{65}{2016}{1-57}

\bibitem{Fre+84}
\Name{Freidlin, M. I. and A. D. Wentzell, A. D.}
\Book{Random Perturbations of Dynamical Systems: 
Grundlehren der Mathematischen Wissenschaften} 
\Vol{260}
\Publ{Springer, New York}
\Year{1984}

\bibitem{Tou09}
\Name{Touchette, H.}
%{\sl The large deviation approach to statistical mechanics},
\REVIEW{Phys. Rep.} {478}{2009}{1}

\bibitem{Tan+03}
\Name{Tanase-Nicola, S. and Kurchan, J.}
%{\sl Statistical-mechanical formulation of Lyapunov exponents}
\REVIEW{J. Phys. A: Math. and Gen.}{36}{2003}{10299}


\bibitem{Tai+07}
\Name{Tailleur, J. and Kurchan, J.}
%{\sl Probing rare physical trajectories with Lyapunov weighted dynamics}
\REVIEW{Nature Physics}{3}{2007}{203-7}

\bibitem{Wil+15}
\Name{Wilkinson, M., Guichardaz, R., Pradas, M. and Pumir, A.} 
%{\sl Power-law Distributions in Noisy Dynamical Systems},
\REVIEW{Europhys. Lett.}{111}{2015}{50005}

\bibitem{Don+76}
\Name{Donsker, M. D. and Varadhan, S. R. S.}
%{\sl Asymptotic evaluation of certain Markov process expectations for large time, I.}
\REVIEW{Commun. Pure Appl. Math.}{28}{1976}{1-47}  
\REVIEW{Commun. Pure Appl. Math.}{28}{1976}{279-301}
\REVIEW{Commun. Pure Appl. Math.}{29}{1976}{389-461}

\bibitem{Wil+03}
\Name{Wilkinson, M. and Mehlig, B.}
%{\sl The Path-Coalescence Transition and its Applications},
\REVIEW{Phys. Rev. E}{68}{2003}{040101}

\bibitem{Hea62}
\Name{Heading, J.}
\Book{An Introduction io Phase Integral Methods}
\Publ{Methuen, London}
\Year{1962}

\bibitem{Lan+58}
\Name{Landau, L. D. and Lifshitz, E. M.} 
\Book{Quantum Mechanics}
\Publ{Pergamon, Oxford}
\Year{1958}

\bibitem{Mil+53}
\Name{Miller, S. C. Jr. and Good, R. H. Jr.}
\REVIEW{Phys. Rev.}{91}{1953}{174-9} 

\bibitem{Fyo+17}
\Name{Fyodorov, Y. V., Le Doussal, P., Rosso, A. and Texier, C.}
arXiv:1703.10066.

\bibitem{Wil+12}
\Name{Wilkinson, M., Mehlig, B., Gustavsson, K. and Werner, E.}
%{\sl Clustering of Exponentially Separating Trajectories},
\REVIEW{Eur. Phys. J. B}{85}{2012}{18}

\bibitem{Som+93}
\Name{Sommerer, J. and E. Ott, E.}
%{\sl Particles floating on a random flow: a dynamically comprehensible physical fractal},
\REVIEW{Science}{359}{1993}{334}

\bibitem{Lar+09}
\Name{Larkin, J., Bandi, M. M., Pumir, A. and Goldburg, W. I.}
%{\sl Power-law distributions of particle concentration in free-surface flows},
\REVIEW{Phys. Rev. E}{80}{2009} {066301}

\bibitem{GC95} 
\Name{Gallavotti, G. and Cohen, E. G. D.}
%{\sl Dynamical ensembles in nonequilibrium statistical mechanics},
\REVIEW{Phys. Rev. Lett.}{74}{1995}{2694}

\bibitem{Sei12} 
\Name{Seifert, U.}
%{\sl Stochastic thermodynamics, fluctuation theorems and molecular machines},
\REVIEW{Rep. Prog. Phys.}{75}{2012}{126001}

\end{thebibliography}
\end{document}